\documentclass{article}
\usepackage{emulateapj}
\usepackage{psfig}
\newcommand{\bo}{\mbox{\boldmath $\mathbf{\omega}$}}

\lefthead{Gu}
\righthead{Vorticity Budget of Weak Thermal Convection in Keplerian disks} 
\submitted{Accepted to ApJ 2000 April 19}
\begin{document}

\def\etal{{\it et al.~}}
\def\eg{{\it e.g.,~}}
\def\ie{{\it i.e.,~}}

\title{Vorticity Budget of Weak Thermal Convection in Keplerian disks} 

\author{Pin-Gao Gu}
\affil{Department of Physics, University of Texas at Austin,
Austin, TX 78712; gu@physics.utexas.edu}
\affil{Department of Physics and Astronomy,
Johns Hopkins University, Baltimore, MD 21218}

\begin{abstract}

By employing the equations of mean-square vorticity
(enstrophy) fluctuations in strong shear flows,
we demonstrate that unlike energy production of turbulent
vorticity in non-rotating shear flows, the
turbulent vorticity of weak convection
in Keplerian
disks cannot gain energy from vortex stretching/tilting by
background shear unless the associated Reynolds stresses
are negative. This is because the epicyclic motion is
an energy sink of the radial component of
mean-square turbulent vorticity in Keplerian disks when
Reynolds stresses are positive. Consequently, weak convection
cannot be self-sustained
in Keplerian flows. This agrees with the results
implied from the equations of mean-square velocity
fluctuations in strong shear flows. Our analysis also sheds light
on the explanation of the simulation result in which
positive kinetic helicity is produced by
the Balbus-Hawley instability in a vertically stratified
Keplerian disk. We also comment on
the possibility of outward angular momentum transport by
strong convection based on azimuthal pressure perturbations
and directions of energy cascade.

\end{abstract}



\section {Introduction}
Turbulence is usually driven by waves or instabilities in
fluids, taping accessible free energy into fluctuating
velocity fields. For astronomical objects,
accessible free energy can come from
body forces (\eg gravity or Lorentz force)
or from large-scale fluid motions (\eg rotation or
shear).

In Keplerian disks, thermal convection is an important ingredient
in the thermal-viscous instability model which explains
the semi-periodic changes of
light curves of dwarf novae (\cite{mm81})
and soft X-ray transients (\cite{mw89}).
However, it has been argued that
convective instability cannot access free
thermal energy (\ie higher entropy near the mid-plane)
to transport energy vertically outward
if it can access free rotational energy (\ie higher
angular velocities at small radii) to transport angular
momentum radially outward.
Recently a more general argument has been
established to include the turbulence which is not thermally
driven and whose azimuthal pressure perturbation is small.
The equations of averaged turbulent kinetic energy
due to local hydrodynamic mixing can be expressed 
as (\cite{bh98})
\begin{eqnarray}
\lefteqn{
\frac{\partial}{\partial t} \left< \frac{\rho v_r^2}{2}
\right> + \nabla \cdot \left< \frac{1}{2} \rho v_r^2 \bf{v}
\right> } \nonumber \\
& & = 2\Omega \left< \rho v_r v_{\theta} \right>
-\left<v_r \frac{\partial \delta P}
{\partial r} \right> - {\rm losses},
\label{vel_r}
\end{eqnarray}

\begin{eqnarray}
\lefteqn{
\frac{\partial}{\partial t} \left< \frac{\rho v_{\theta}^2}{2}
\right> + \nabla \cdot \left< \frac{1}{2} \rho v_{\theta}^2 \bf{v}
\right> } \nonumber \\
& & = -\frac{\kappa^2}{2\Omega}
\left< \rho v_r v_{\theta} \right>-\left<v_{\theta} 
\frac{\partial \delta P}
{r\partial \theta} \right> - {\rm losses},
\label{vel_theta}
\end{eqnarray}
where ``losses'' represent the energy sink due to viscosity,
$v_i$ is the turbulent velocity field, $\delta P$ is the pressure
perturbation,
$\rho$ is the mass density, $\Omega$ is the Keplerian angular
velocity,
and the epicyclic frequency $\kappa^2=(2\Omega/r)d(r^2\Omega)/dr$.
The notation $< \ >$ in equations (\ref{vel_r}) and
(\ref{vel_theta}) denotes the averaging
carried out over the ensemble of turbulent cells, indicating
the time/space correlation between two fluctuating quantities.
In the case of accretion disks, the time average is taken
over several eddy turnover time (but still smaller than the
viscous time scale), and the space average is performed
over $2\pi$ in the azimuthal direction, over the whole disk vertical
scale height\footnote{Averaging over the disk height is not
performed when one would like to investigate the detailed vertical
structures (\eg see \cite{SHGB96}; \cite{ms00}.)}, 
and over several eddy sizes in the radial direction
(but still much smaller than the disk radius).

As indicated by equation (\ref{vel_theta}),
the epicyclic term is an energy sink of $\left< v_{\theta}^2 \right>$
if $\left< v_r v_{\theta}
\right> >0$.
Therefore any local hydrodynamic instabilities
cannot grow when azimuthal pressure perturbation is small.
If a disk, however, is heated
by other sources, the energy of weak convection
can be maintained and then transport angular momentum inward.
When thermal convection is weak,
the associated negative Reynolds stress is a result of
conservation of angular momentum of turbulent
elements during
the process of local mixing in Rayleigh-stable disks
such as Keplerian flows (\cite{balbus00}).

Besides considering the epicyclic effect which damps 
fluctuating velocity fields,
the same damping effect should apply to
fluctuating vorticity
fields which characterize strong turbulence. It is also
worth investigating how fluctuating vorticity fields interact
with background vorticity when convection is maintained by
other heating sources such as MHD turbulence driven by
the Balbus-Hawley instability (\cite{bh98}),
in order to determine
the typical linear modes for convection
in nonlinear regime (\cite{gvc}, hereafter
GVC). As high levels
of fluctuating vorticities are observed in non-rotating shear
flows at high Reynolds numbers, we must be wondering why the
generation of fluctuating vorticities in Rayleigh-stable disks
cannot be achieved by the usual vortex dynamics such as
vortex stretching or vortex tilting, as noted and
numerically simulated by \cite{hbw99}.

Perhaps one of
the most difficult tasks in dealing with equations of turbulence
is to determine the signs of correlations between fluctuating
quantities. Positive, negative, or no correlations are usually
related to the properties of background
flows such as stratification, rotation,
background shear, or body forces.
Positive Reynolds stresses in non-rotating shear flows, 
for example, are always related to
the mean flow moving in $+x$ direction with negative gradient in
$+y$ direction. 
This situation is reversed in Keplerian flows for weak
convection (\cite{rg92}; \cite{kpl93}; \cite{knl95};
\cite{cab96}; \cite{sb96}; \cite{balbus00}).
Positive kinetic helicity occurring in magnetized disks
shown in the numerical simulation by \cite{bd97} probably
means that the Coriolis force is of less importance in
Keplerian disks where MHD turbulence is driven by the
Balbus-Hawley instability.
Similar to these examples,
we would expect that background shear and rotation
can determine the signs of correlations
in the equations of fluctuating vorticities.

Beginning with the perturbed vorticity equations in a non-rotating
flow, we determine the signs of turbulent correlation
based on positive Reynolds stresses without worrying
about nonlinear shear instabilities in \S 2. In \S 3, we apply
the analysis presented in \S 2 to Keplerian flows, showing
that mean-square vorticity perturbations of
weak convection cannot grow with positive Reynolds stresses.
Positive kinetic helicity driven by the Balbus-Hawley
instability in a stratified Keplerian disk is analyzed
through vortex equations in \S 4.
In the last section,
we comment on the possibility of outward angular momentum by
strong convection
in accretion disks based on the recent development of theories
concerning about azimuthal pressure perturbations and directions
of energy cascade.

\section {momentum conservation and vorticity equations in non-rotating
shear flows}
The vorticity equation reads
\begin{equation}
\frac{D {\bf w}}{Dt}=\left( {\bf w} \cdot \nabla \right) {\bf u}
-\left( \nabla \cdot {\bf u} \right) {\bf w}
+\frac{\nabla \rho \times \nabla P}{\rho^2}+\nabla \times \nu \nabla^2
{\bf u},
\label{vor}
\end{equation}
where ${\bf w}=\nabla \times {\bf u}$,
$D/Dt=\partial_t + ({\bf u} \cdot \nabla)$,
$P$ is the pressure,
$\rho$ is the mass density, and $\nu$ is the kinematic viscosity.
Consider a 3-D turbulent flow with a background shear
$dV/dx$ without rotation, where $V(x)$ is the background flow
in $+y$ direction. Let the vertical scale height be much smaller
than the scale heights in $x$ and $y$ directions.
If the eddy growth rate is smaller than the shearing
rate (\ie weak convection),
linearizing the vorticity equation (${\bf u}=V{\bf \hat \jmath}+{\bf v}$,
and $\bo = \nabla \times {\bf v}$)
gives
the equations of mean-square vorticity (enstrophy) perturbation:
\begin{eqnarray}
\lefteqn{\frac{D}{Dt} \left< {\omega_x^2 \over 2} \right> \approx 
\left<  \omega_x \frac{\partial v_x}{\partial z} 
\right> \frac{dV}{dx}} \nonumber \\
& & -{1\over \rho^2}{\partial \rho \over \partial z}\left< \omega_x
{\partial \delta P \over \partial y} \right>
+{1\over \rho^2}{\partial P \over \partial z}\left< \omega_x
{\partial \delta \rho \over \partial y} \right>
 -{\rm losses},
\label{vor_x}
\end{eqnarray}

\begin{eqnarray}
\lefteqn{
\frac{D}{Dt}  \left<  {\omega_y^2 \over 2} \right>
\approx  \left< \omega_x \omega_y \right> \frac{dV}{dx}
+\left< \omega_y \frac{\partial v_y}{\partial z} 
\right> \frac{dV}{dx} } \nonumber \\
& & +{1\over \rho^2}{\partial \rho \over \partial z}\left< \omega_y
{\partial \delta P \over \partial x} \right>
- {1\over \rho^2}{\partial P \over \partial z}\left< \omega_y
{\partial \delta \rho \over \partial x} \right>
-{\rm losses},
\label{vor_y}
\end{eqnarray}

\begin{eqnarray}
\lefteqn{
\frac{D}{Dt} \left< {\omega_z^2 \over 2} \right>
\approx -\left< v_x \omega_z \right> \frac{d^2V}{dx^2} } \nonumber \\
& & - \left< \omega_z \left( 
\frac{\partial v_x}{\partial x} +\frac{\partial v_y}{\partial y}
\right) \right> \frac{dV}{dx} -{\rm losses},
\label{vor_z}
\end{eqnarray}
where the terms ``losses'' represent the energy sink due to
viscosity, $v_i$ and $\omega_i$ are
fluctuating velocity and vorticity respectively,
and $D/Dt$ denotes $\partial_t + V\partial_y$.
We have ignored the term $\left< \omega_i \omega_j \partial_j v_i \right>$
which represents stochastic stretching of vortex by
turbulent shear.
This effect is smaller than stretching
by background shear as long as
background shear rate is larger than turbulent growth rate.
Similar to the role played by $\partial \delta P/\partial r$
in equation (\ref{vel_r}), the terms associated with
$\partial \delta P/\partial x$ and $\partial \delta \rho/\partial x$
are energy sources of $\left< \omega_y^2 \right>$ in equation
(\ref{vor_y}). Since these terms come from the baroclinic
term in equation (\ref{vor}), they appear as a result of
vertical stratification.

The production or annihilation of turbulent vorticity
relies on pressure perturbations, density perturbations, and
the signs of correlations between fluctuating
vorticities and fluctuating velocity shears as shown in
the above equations. However, the term
associated with $\partial_y$ is usually small
for weak convection as a result of strong shear.
Assuming that all perturbation quantities are proportional
to $\exp (ik_x x+ik_y y+i k_z z+i\omega t)$, we have
the linear perturbation equations for 
adiabatic convection under Boussinesq approximation:
\begin{eqnarray}
i\bar \omega v_x +ik_x \Psi &=&0,
\label{per2}\\
i\bar \omega v_{y}+{dV \over dx} v_x +ik_y \Psi &=&0,
\label{per3}\\
i\bar\omega v_z + ik_z\Psi +\delta g_z &=&0,\label{per4}\\
i\bar\omega\delta-v_z \partial_z
\ln\left({P^{1/\Gamma}\over\rho}\right)
&=&0,
\label{per5}\\
k_xv_x+k_y v_y+k_zv_z&=&0,\label{per6}
\end{eqnarray}
with the shearing constraint 
(\cite{shu74})\footnote{This constraint is called
the ``weak-shear limit'' in \cite{shu74}. In fact, we are
dealing with convection with the growth rates smaller
than the shearing rate in this paper. In order to avoid
confusion, we do not use the term ``weak-shear''
here.}
\begin{equation}
k_y < k_x {i\bar \omega \over |dV/dx|},\label{per7}
\end{equation}
where $\bar \omega$ is the frequency measured by a local
observer comoving with the mean flow, $\delta\equiv \delta \rho
/\rho$ is the local fractional density perturbation,
$\Psi\equiv \delta P/\rho$ is the pressure perturbation
divided by the density, $v_x$, $v_y$, and $v_z$ are
velocity perturbations, $k_x$, $k_y$ and $k_z$ are the
wavenumbers in $x$, $y$ and $z$ directions.
The validity of the plane-wave approximation for
$k_x$ is assured by the shearing constraint
for thermal convection in the presence of a shear flow.
The plane-wave approximation for $k_z$ in the presence
of a vertical gradient is not a bad assumption for
weak convection since first of all, the dispersion
relation (\ref{disper})
inferred from the above Boussinesq equations in
the case of accretion disks is the same as the
one derived by \cite{rpl88}, who consider the vertical
wavelength of axisymmetrical modes for convection
as a function of $z$ owing to vertical stratification.
Second of all, as we shall see in the following text,
the equations (\ref{per2})--(\ref{per7})
serve only as an example to illustrate how the phenomenological
interpretation works for explaining the directions of
angular momentum transport done by \cite{balbus00}, and
how convective eddies approximately look like in a shear flow,
all of which are not very sensitive to the detailed vertical
structures of disks. We are going to adopt the same phenomenological
strategy to evaluate the signs of correlations in the equations
of perturbed enstrophy, and consequently to investigate if turbulent
vorticities can be generated by vortex stretching/tilting
in the regime where the terms associated with $\partial_y$ are small.

By virtue of equations (\ref{per2}) through (\ref{per7}) in the
weak-convection limit (\ie the convective growth rate
$i \bar \omega < |dV/dx|$),
one can show that the terms associated with $\partial_y$
(or say $k_{\theta}$)
in the above Boussinesq equations and in the equations of
perturbed enstrophy are smaller than
the terms associated with $dV/dx$,
owing to the shearing constraint. Without the pressure
perturbation in the $y$ direction, equation (\ref{per3})
states that a turbulent element can
conserve momentum in the $y$ direction when it moves.

In the case $dV/dx<0$, conservation
of momentum of turbulent elements in the $y$ direction
described by equation (\ref{per3})
implies that
$\left<v_x v_y \right> > 0$, \ie as observed in the comoving
frame, positive (negative) values
of $v_x$ should occur more frequently than
negative (positive) ones when $v_y$ is positive (negative),
or vice versa. By virtue of equation (\ref{per3}), $\left< v_y^2
\right>$ can grow when $\left< v_x v_y \right> >0$ (\cite{bh98}).
The major concern for a successful
growth of turbulent vorticity is that
the first term of the right hand side in equation (\ref{vor_x})
must be positive because the other terms associated with
$\partial_y$ are small. Since it is the equation
(\ref{per3}) which determines the sign of $\left< v_x
v_y \right>$ and the growth of turbulence, 
we would expect that the same equation plays the same role
in equation (\ref{vor_x}).
By virtue of
equation (\ref{per3}) again, we have
\begin{equation}
{1\over 2} \left< {\partial \gamma \over \partial z}
{\partial v_y^2 \over \partial z} \right> +
\gamma \left< \left( {\partial v_y \over \partial z} \right)^2 \right>
=-{dV \over dx} \left< {\partial v_y \over \partial z}
{\partial v_x \over \partial z} \right>,
\end{equation}
where the growth rate $\gamma \equiv i\bar \omega >0$. The
first term on the left hand is positive since turbulent
velocity fields usually increase with convective growth rate. As
a result, the correlation $\left< \partial_z v_y
\partial_z v_x \right>$ is positive.
The first term
on the right hand side of equation (\ref{vor_x}) is
$\approx -\left< \partial_z v_y \partial_z v_x \right> dV/dx$
which is therefore positive, leading to successful growth
of $\left< \omega_x^2 \right>$.
In fact, this positive sign can be
understood in another way. The equation of vorticity
$D\omega_x /Dt \approx \partial_z v_x (dV/dx)$ suggests that
$\omega_x$ tends to be negative (positive) when
$\partial_z v_x >0$ ($<0$) and $dV/dx<0$ ($>0$). Hence
$\omega_x$ and $\partial_z v_x$ are negatively correlated
in a non-rotating shear flow.
In other words, $\left< \omega_x^2 \right>$
can grow from the interaction between background ($dV/dx$)
and turbulent ($\partial_z v_x$) shear.

Although the signs of correlations associated with background
shear in equation (\ref{vor_y})
are not important because there are
energy source terms due to pressure/density perturbations,
we can possibly determine the sign of $\left< \omega_x
\omega_y \right>$ phenomenologically. In the case of
$dV/dx <0$, if we consider a turbulent element
moving toward $+x$ direction with positive sense of $\omega_y$,
the element will tend to move toward $+y$ direction more
frequently so that the negative sense of $\omega_x$ is created
as seen in the comoving frame. This effect of vortex rotation
is therefore described by the correlation
$\left< \omega_x \omega_y \right> <0$.
The signs is reversed when $dV/dx>0$. After all, the
term $\left< \omega_x \omega_y \right> dV/dx$ is an energy source
of $\left< \omega_y^2 \right>$.

The term $\left< v_x \omega_z \right> $ in equation (\ref{vor_z})
describes the vorticity transport in the flows which have nonuniform
background vorticity (\ie $d^2V/dx^2 \neq 0$).
Assume that the angular momentum of vertical
vortex tubes are nearly conserved; \ie the loss term in
equation (\ref{vor_z}) is small.
If the gradient of background vorticity
$d^2V/dx^2$ is positive (negative), the local turbulent mixing will
transport the vorticity toward $-x$ ($+x$) direction, or vice versa.
This means that the term
$-\left< v_x \omega_z \right> d^2V/dx^2 >0$. However, the production
provided by vorticity transport is supposed
to be small since $v_x$ and the correlation length across the
mean shear flows are severely shorten by strong shear. This
point can be also understood in terms of the fact that
the term $\left< v_x \omega_z \right>$ is zero in a
shear flow without a gradient of a background state in
the $x$ direction. Introducing a gradient of background
state gives rise to a factor about one over the scale height
along the $x$ direction which is much smaller than one over
the vertical scale height associated with other terms in
equation (\ref{vor_z}).

The second term in equation (\ref{vor_z}) describes vortex
generation due to the interaction between
background vorticity $dV/dx$ and variation of cross sections
of vortices.
The equation of motion
$D\omega_z /Dt \approx -(\partial_x v_x+\partial_y v_y)(dV/dx)$ suggests
that $\omega_z$ tends to be negative (positive) when
$\partial_x v_x+\partial_y v_y<0$ ($>0$)
and $dV/dx <0$ ($>0$).
This means that the second term in
equation (\ref{vor_z}) is the energy source of
$\left< \omega_z^2 \right>$. For a nearly incompressible
disturbance, $\partial_x v_x+\partial_y v_y<0$ means vortex
stretching (\ie $\partial_z v_z>0$), and
$\partial_x v_x+\partial_y v_y>0$ means vortex
squeezing (\ie $\partial_z v_z <0$) in the vertical direction.

After all, the turbulent vorticities $\omega_x$ and $\omega_y$
can grow simultaneously via
vortex tilting by background
shear ($dV/dx$) and by turbulent shear ($\partial_j v_i$).
The turbulent vorticity $\omega_z$ can grow via vortex
stretching/squeezing. The different vortex dynamics associated
with $\omega_z$ comes from the fact that
$\omega_z$ does not couple with $\omega_x$ and $\omega_y$
as shown in equations (\ref{vor_x}), (\ref{vor_y}), and
(\ref{vor_z}).
We note that vortex stretching/tilting is a 3-D
effect and cannot exist in 2-D fluids ($\omega_x \approx \omega_y
\approx 0$)
which are usually employed
to model vortex generation in larger scales $>>$ vertical
scale height $h$ 
(\ie
the shallow water approximation). Large-scale vortex production in
2-D usually relies on the baroclinic term in the first place,
and subsequently vortexes
can evolve by mutual interactions such as merging into large
ones which have long lifetimes (\cite{aw95}; \cite{gl99}).
In this paper, we do not consider this evolution.

Before we apply the same approach to Keplerian flows,
we need to emphasize that the above analysis just serves
as preliminary calculations for Keplerian flows.
The turbulent fields driven by thermal convection in a
planar shear flow are more complicated than those in
a Keplerian flow. In the former case, thermal convection
excites a non-linear shear instability which
overwhelms the convective motion (\cite{sb96}), 
and therefore the
above analysis for thermal convection in a planar
shear flow is quite questionable; for example, the term
$\left< \omega_i \omega_j \partial_j v_i \right>$
which we ignored becomes most important (\cite{tl72}). 
In the case of Keplerian flows,
however, numerical simulations thus far have shown that
strong epicyclic motion suppresses
non-linear shear instabilities 
(\cite{hbw99}, however see \cite{rz99},
and \cite{kla00}). 
Hence the above analysis
should be reasonable for weak convection in
a Keplerian flow. We shall
see in the next section that the result implied from
the equations of mean-square vorticity perturbations
are indeed consistent with the result inferred from
the equations of mean-square velocity fluctuations.

\section{angular momentum conservation and
vorticity equations in a Keplerian flow}
In Keplerian flows, the vorticity equations
in cylindrical coordinates
become
\begin{eqnarray}
\lefteqn{
\frac{D}{Dt} \left< {\omega_r^2 \over 2} \right>
\approx
\left< \omega_r \frac{\partial v_r}{\partial z} \right> \frac{\kappa^2}
{2\Omega} } \nonumber \\
& & -{1\over \rho^2}{\partial \rho \over \partial z}\left< \omega_r
{\partial \delta P \over r\partial \theta} \right>
-{\rm losses},
\label{vor_r}
\end{eqnarray}

\begin{eqnarray}
\lefteqn{
\frac{D}{Dt} 
\left< {\omega_{\theta}^2 \over 2} \right> 
\approx
\left< \omega_r \omega_{\theta} \right> r\frac{d\Omega}{dr}
} \nonumber \\
& & +\left< \omega_{\theta} 
\frac{\partial v_{\theta}}{\partial z} \right> \frac{\kappa^2}
{2\Omega} 
+{1\over \rho^2}{\partial \rho \over \partial z}\left< \omega_\theta
{\partial \delta P \over \partial r} \right>
-{\rm losses},
\label{vor_theta}
\end{eqnarray}

\begin{eqnarray}
\lefteqn{
\frac{D}{Dt} \left< {\omega_z^2\over 2} \right>
\approx
-\left< v_r \omega_z \right> \frac{d}{dr}\left(
\frac{\kappa^2}{2\Omega} \right)
} \nonumber \\
& & -\left< \omega_z \left( {\partial v_r \over \partial r}
+{\partial v_{\theta} \over r\partial \theta} \right) \right> \frac{\kappa^2}
{2\Omega} -{\rm losses},
\label{vor_zz}
\end{eqnarray}
where $D/Dt$ denotes
$\partial_t +\Omega \partial_{\theta}$. In a
Keplerian disk threaded with sub-thermal magnetic fields,
the ``loss'' terms in above equations include the turbulent
damping due to radial mixing driven by the
Balbus-Hawley instability. We do not show the terms
associated with density perturbations explicitly because
unlike $\delta P/P \lesssim \delta \rho /\rho$ in non-rotating
shear flows owing to equation (\ref{per2}),
 $\delta P/P \sim \delta \rho /\rho$ in rotating shear flows
due to the Coriolis force (see equation (\ref{v_r}) below).
Once again $\omega_z$ does not couple with $\omega_r$ and
$\omega_{\theta}$.

Although perturbation equations (\ref{per2}) through (\ref{per7})
should be modified by adding the Coriolis force in a rotating flow,
the Coriolis term $2\Omega$
is comparable to the shear term $rd\Omega /dr$. Therefore
the terms associated with  $\partial_{\theta}$
are small compared to the first term on the right hand side
in equation (\ref{vor_r}),
analogous to equation (\ref{vor_x}).
In a Keplerian flow, neglecting
azimuthal pressure perturbations means that turbulent
elements conserve
angular momenta before they mix
with background fluid. This leads to a transport
of angular momentum down to the angular momentum
gradient, giving rise to negative Reynolds stresses
in a Keplerian disk (\cite{balbus00}). This picture is
described by the azimuthal perturbed equation of motion
which is similar to equation (\ref{per3}) except dV/dx is
replaced by $\kappa^2/2\Omega$; \ie
\begin{equation}
\gamma v_{\theta} =-{\kappa^2 \over 2\Omega }v_r
=-\left( r{d\Omega \over r}+2\Omega \right) v_r,
\label{v_theta}
\end{equation}
where we have dropped the term $ik_{\theta} \Psi$.
The above equation indicates that $\left< v_{\theta}^2 \right>$
cannot grow when $\left< v_r v_{\theta} \right> >0$
(\cite{bh98}), which is different from the situation in a
non-rotating shear flow owing to the appearance of the Coriolis force.

In fact, negative Reynolds stresses can be realized as follows.
Consider a turbulent element is moving with positive $v_r$. While
shear ($rd\Omega/dr$) tries to move the element to the $+\hat \theta$,
as observed in a corotating frame, the
Coriolis force ($2\Omega$) tried to move the element to the
$-\hat \theta$. Since $2\Omega > r|d\Omega/dr|$ in Keplerian disks
(\ie $\kappa^2 >0$), the Coriolis force is the winner.
Therefore the turbulent
element tends to have $-v_{\theta}$, leading to
$\left< v_r v_{\theta} \right> <0$. Without rotation (Coriolis effect),
$\left< v_x v_y \right> >0$ in a non-rotating
shear flow with negative $dV/dx$. 
The contribution to $\left< v_r v_{\theta} \right>$
from the radial deviation of
turbulent azimuthal motion due to the Coriolis force
is small since
for weak convection in a Keplerian disk, equation (\ref{per2})
becomes
\begin{equation}
ik_r \Psi \sim 2\Omega v_{\theta} \label{v_r},
\end{equation}
where the term $i\bar \omega v_r$ is small and has been ignored.
This means that the strong radial gradient of pressure perturbation
is roughly balanced by the Coriolis force, resulting in
negligible contribution to $\left< v_r v_{\theta} \right>$.
In other words,
$\left< v_r v_{\theta} \right>$ is mostly determined by equation
(\ref{v_theta}). We note that in contrast to equation
(\ref{per2}), the extra term due to the Coriolis force
in equation (\ref{v_r}) indicates that the radial
gradient of pressure perturbation should be large enough
to make convection grow in a rotating flow.
Namely, the Coriolis force is a stabilized
factor to convection. Owing to equations (\ref{v_theta})
and (\ref{v_r}), the linear theory without radiative
and turbulent damping indicates that $\gamma^2$ is smaller
than the Brunt-V\"ais\"al\"a frequency $-N^2$
(GVC; Ruden \etal 1988):
\begin{equation}
\gamma^2={-N^2 -A \Omega^2 \over A+1},
\label{disper}
\end{equation}
where $A\equiv (k_z/k_r)^2$.
It has been known in
terrestrial experiments that the critical Rayleigh
numbers of rotating Rayleigh-Benard convection
are increased above their non-rotating 
values (\eg \cite{zes93}).

The above analysis means that for weak convection,
the signs of
correlations in Keplerian flows in equations (\ref{vor_r})
(\ref{vor_theta}), and (\ref{vor_zz})
are reversed compared with
those in the non-rotating flows with $dV/dx<0$ when we replace
$x-y$ with $r-\theta$.
In other words, we have
$\left< v_r v_{\theta} \right> < 0$,
$\left< \omega_r \omega_{\theta} \right> >0$,
$\left< \omega_r \partial_z v_r \right> >0$, and
$\left< \omega_z (\partial_r v_r+ 
\partial_{\theta} v_{\theta}/r) \right> <0$
when $\kappa^2>0$.
The right hand side of equation (\ref{vor_r}) becomes positive
and $\left< \omega_r^2 \right>$ can grow. Although the first
term on
the right hand side of equation (\ref{vor_theta}) turns out to be
negative, $\left< \omega_{\theta}^2 \right>$ can be maintained
by the radial gradient of perturbed pressure owing to
strong epicyclic effect.

If we assume that weak convection
transports angular momentum down to the angular velocity
gradient, then the signs of correlations in equation (\ref{vor_r})
and equation (\ref{vor_theta}) are the same as those in the
non-rotating flows with $dV/dx<0$ when we replace $x-y$
with $r-\theta$. This means that the first term
on the right hand of equation (\ref{vor_r}) is negative
because $\kappa^2>0$. Hence $\left< \omega_r^2 \right>$
cannot grow and weak convection dies away.

If the disks are Rayleigh-unstable (\ie $\kappa^2 <0$),
the signs of all terms in equations (\ref{vor_r}), (\ref{vor_theta}),
and (\ref{vor_zz}) remain unchanged compared to those
with positive $\kappa^2$
except that the first
term on the right hand side in equation (\ref{vor_theta})
becomes positive because
$\left< \omega_r \omega_{\theta} \right> <0$. In other words,
a Rayleigh-unstable disk behaves as a non-rotating shear flow
in the sense that
$\left< v_r v_{\theta} \right> >0$.
We note that
our analysis for Rayleigh-unstable disks is oversimplified
because a non-linear shear instability should be excited.

\section{positive kinetic helicity as a result of the
Balbus-Hawley instability}
A weakly magnetized Keplerian disk is linearly 
unstable to the Balbus-Hawley instability. In nonlinear
regime, this
instability can drive MHD turbulence which in
turn leads to a dynamo process unless the magnetic
Reynolds numbers are low (\cite{bh98}). 
\cite{bd97} shows that this
dynamo process observed in simulations
can be imitated by an $\alpha -\Omega$ dynamo
in a vertically stratified disk,
with a negative $\alpha_{\theta \theta}$ in the upper disk
plane (where $\alpha_{\theta \theta}$ is the
$\theta$-$\theta$ component of kinetic helicity tensor $\alpha$).
A negative $\alpha_{\theta\theta}$ means that
the kinetic helicity $\left< v_z \omega_z \right>$
is positive for nearly isotropic turbulence.
This contradicts with the usual notion that the
kinetic helicity of convection driven by the Coriolis force
is usually negative. \cite{bran98} suggests that the right-handed
helical turbulence results from the combined effect of
the Balbus-Hawley instability (\ie $\left< \delta B_r
\delta B_{\theta} \right> <0$) and magnetic buoyancy.

Although we have concentrated on convection-like eddies and
have not taken into account the effect of
sub-thermal magnetic fields in this paper, equations
(\ref{vor_z}) and (\ref{vor_zz}) could shed light on
preferred directions of helical turbulence
if we realize that a Coriolis-dominated disk
characterized by equation (\ref{vor_zz}) becomes
a shear-dominated disk symbolized by equation
(\ref{vor_z}) due to the Balbus-Hawley instability.
As explained in the preceding section,
$\left< \omega_z (\partial_r v_r+
\partial_{\theta} v_{\theta}/r) \right> <0$
in a nonmagnetized Keplerian disk. 
Assume that
a turbulent element 
expands (contracts) as it goes upward (downward)
in the upper disk plane. The above negative
correlation means that a turbulent
element tends to have a negative (positive)
$\omega_z$ due to expansion (contraction). Consequently,
$\left< v_z \omega_z \right> <0$ (see the vortex plot
associated with positive $\kappa^2$ in Figure 1).
This result is
consistent with the picture in which the Coriolis force
is the winner over shear so that turbulent eddies
are mostly left-handed; \ie as observed in a corotating frame,
left-handed helical turbulent motion results from a positive
vorticity ($\kappa^2 >0$) of the mean flow. However, introducing
the Balbus-Hawley
instability changes the sign of Reynolds stresses
from negative to positive in Keplerian disks because
angular momentum of a turbulent element is changed by
fluctuating magnetic torques (\cite{balbus00}), leading
to the condition for the instability changed from 
the Rayleigh criterion $\kappa^2<0$ 
to the Chandrasekhar-Velikov criterion
$d\Omega^2/dr <0$ (\cite{vel59}; \cite{chan60}; \cite{fri69})
in a Keplerian flow. Equivalently speaking,
the Velikov-Chandrasekhar criterion
is a manifestation of conservation of angular velocity of
turbulent elements due to the fact that the fluctuating magnetic
torque counteracts the torque of
the Coriolis force in a weak field limit (\cite{fri69}).
Without the torque by
the Coriolis force, azimuthal dynamics of turbulence in
Keplerian disks becomes that in a non-rotating shear flow.
Furthermore,
if the contribution to $\left< v_r v_{\theta}
\right>$ due to the radial deviation of azimuthal
motion is small (like the situation described by
equation (\ref{v_r}); radial deviation of azimuthal
motion is usually small owing to shear),
then the signs of correlation quantities
in a weakly magnetized disk should
behave as the one described by negative vorticity
of the mean flow (such as the correlation quantities in
equations (\ref{vor_x}), (\ref{vor_y}), and (\ref{vor_z})), 
giving rise to
a positive kinetic helicity (see the vortex picture associated
with negative $dV/dx$ in Figure 1). 
The only physical difference
of vortex dynamics between a non-rotating shear flow and
a weakly magnetized disk
is that in the later case,
a positive kinetic helicity coupled with a positive
vorticity of the mean flow becomes an energy sink of
$\left< \omega_z^2 \right>$ as shown in equation (\ref{vor_zz}).
This is analogous to the result that a positive Reynolds
stress is an energy sink of $v_{\theta}^2$ in Keplerian disks
(\cite{hbw99}).

\hbox{~}
\centerline{\psfig{file=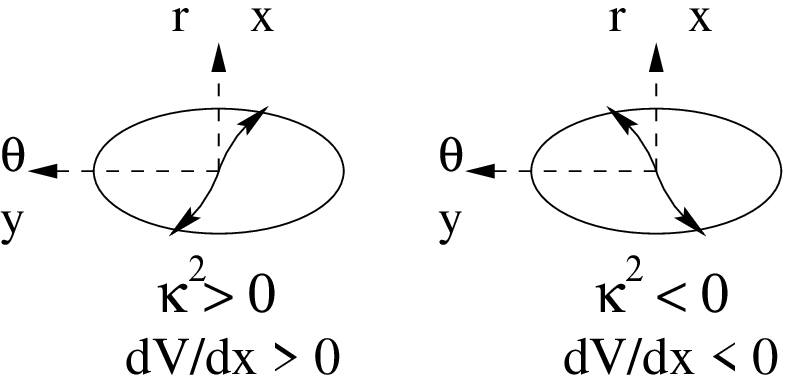,angle=0,width=3in}}
\vspace{0.1in}
\noindent{\small
\addtolength{\baselineskip}{13pt} 
\hspace*{0.3cm} Fig.~1.\
The expansion of a rising turbulent element
described by equation (\ref{vor_zz}) or by equation (\ref{vor_z})
when $\kappa^2>0$ or $dV/dx >0$
(left), and when $\kappa^2<0$  or $dV/dx <0$ (right).
As observed in a corotating frame, 
$\left< \omega_z (\partial_r v_r +
\partial_{\theta} v_{\theta}/r)  \right> <0$
and $\left< v_z \omega_z \right> <0$ when
$\kappa^2>0$. These signs are reversed when $\kappa^2<0$.
A negative (positive)
$\omega_z$, as a result of a positive (negative) vorticity of
the mean flow, can give rise to a negative (positive)
$\left< v_r v_{\theta} \right>$ when radial deviation of
azimuthal motion is small. As argued in the text, the
picture on the right
could also apply to the $\omega_z$ produced by the
Balbus-Hawley instability which generates positive Reynolds
stresses in Keplerian disks. In the magnetic case,
however, the
vortex motion illustrated in the figure becomes an energy sink
of $\left< \omega_z^2 \right>$. $k_{\theta}<k_r$ is a result
of shear.
\vspace{0.1in}
\addtolength{\baselineskip}{13pt}
}

Although our analysis is related to
convection-like eddies (\ie $k_z \sim 1/h$),
the picture we present here
for estimating the sign of kinetic helicity
does not contradict the
result that the Parker instability, a magnetohydrodynamic
instability with  $k_z \sim 1/h$, is suppressed
by MHD turbulence (\cite{vd92}). The main
point presented here depends solely on the sign
of Reynolds stresses which is assumed to be
largely caused by the azimuthal deviation
of radial motion. Therefore, the picture that we sketch
here applies to any turbulent eddies which expand (contract)
as they rise (descent). We note that rising turbulent eddies
can expand statistically even without
any types of buoyancy. As a matter of
fact, the vertical structures of large-scale magnetic fields
generated by the Balbus-Hawley instability in vertically stratified
accretion disks are stable against magnetic buoyancy 
(\cite{SHGB96}; \cite{ms00}). The detailed information,
such as the magnitude of this kinetic helicity and
the expansion of turbulent eddies in a stably stratified disk,
should depend strongly on the detailed structure of the flow,
such as azimuthal perturbations (\ie anisotropic turbulence)
and correct vertical dynamics.

\section{Discussion and Outlook: comments on angular
momentum transport by strong convection}
In this paper, we demonstrate that
weak convection cannot be self-sustained by vortex
stretching and tilting in a Keplerian flow, 
suggested by the
equations of enstrophy fluctuations. This is because epicyclic
term is an energy sink of $\left< \omega_r^2 \right>$
when $\left<v_r v_{\theta} \right> >0$. The
situation is reversed when weak convection
has a negative Reynolds stress, leading to a successful
growth of thermal convection. Our determination
of the signs of correlation quantities in the equations
of enstrophy is based on
the point of view in which the signs
of Reynolds
stresses are determined by conservation of
angular momentum for weak convection (\cite{balbus00}).
Although our determination of signs of correlation 
quantities is not a rigorous proof but quite suggestive,
the results are consistent with those
suggested by the equations of velocity fluctuations.
Full three-dimensional numerical simulations are
needed to verify our thought.

Positive kinetic helicity in a weakly magnetized, 
vertically stratified Keplerian disk
implied from our analysis agrees with the simulation result by
\cite{bd97}. However, the physical picture shown in Figure 1
is totally different from the plot sketched in Figure 7 in
their paper: in a corotating frame of a Keplerian disk,
the direction of Coriolis effect ($2\Omega$) should be opposite to
the direction of shear effect ($d\Omega/dr$), resulting in a
positive vorticity ($\kappa^2>0$). In our analysis for 
 turbulence, $\left< w_z 
(\partial_r v_r + \partial_{\theta} v_{\theta}/r)
\right> <0$ ($>0$) when vorticity is positive
(negative), giving rise to negative (positive)
$\left< v_z \omega_z \right>$ as long as 
$\left< v_z (\partial_r v_r + \partial_{\theta} v_{\theta}/r)
\right>$ is positive.
Since the Balbus-Hawley
instability generates
a positive $\left< v_r v_{\theta} \right>$ which resembles
the hydrodynamic turbulence with a negative vorticity, MHD
turbulence driven by the Balbus-Hawley instability
should display right-handed helical motion as observed
in numerical simulations. 
However, if rising eddies contract
statistically, our analysis will give rise to the conventional sign
of kinetic helicities in a rotating fluid influenced mostly by
the Coriolis force. The expansion of rising turbulent elements
in a stably stratified disk could be just a result of a
special profile of underlying background states, or
might result from the possible
scenario that
turbulent mixing, an effect definitely not included in usual
laminar analyses, alters the condition of
flux freezing in some degree.
On the contrary, the simple explanation by \cite{bran98} relying
on magnetic buoyancy is not consistent with the simulations
for vertically stratified
accretion disks stirred by the Balbus-Hawley instability.

GVC studied convective energy transport in Keplerian
disks stirred by the Balbus-Hawley instability. The typical
modes of convection are determined from linear perturbation
theory and nonlinear saturation which is caused by the
secondary Kelvin-Helmholtz instability (\ie interaction
between convective shear and background vorticity)
and by MHD turbulent damping.
From equations (\ref{vor_r}), (\ref{vor_theta}),
and (\ref{vor_zz}), we have seen
how convective vorticity/shear possibly interacts with background
vorticity/shear and how this interaction becomes an energy sink
of turbulent vorticity.
This means that the quasi-linear approach for the mixing-length 
theory by GVC based upon the secondary Kelvin-Helmholtz instability
is qualitatively correct.

However, the issue of angular momentum transport by
weak convection is probably not of great importance in reality.
According to the thermal-viscous instability model,
weak convection occurs in Keplerian disks
when the Shakura-Sunyaev viscosity $\alpha_{SS}$
(\cite{ss73})
is large $\sim 0.1$, as a result of
strong radiative losses (GVC; \cite{cab96}).
Furthermore, based on GVC,
large $\alpha_{SS}$
means strong MHD turbulent mixing driven by the
Balbus-Hawley instability, which suppresses weak
convection by smoothing away momentum and entropy
anisotropies associated with convective bubbles mostly in
the radial direction (see equation (\ref{disper}) below):
$A<<1$ when $-N^2<<\Omega$).
Consequently, only the convection
with larger Rossby numbers $Ro \sim 1$ can survive in
a magnetized Keplerian disk. Weak convection can occur
at smaller radii in protostellar disks (\eg see \cite{DAl98}).
In such a cold disk,  a significant `dead zone' implies that the 
Balbus-Hawley instability may not function 
properly, but accretion might occur near the surfaces of
disks where cosmic rays (\cite{gam96}), irradiation
(\cite{gla97}), or rigorous heating beyond the
bottom of photosphere as manifested by protostellar winds
and flares might increase the ionization level beyond the critical 
values (\cite{gm98}).
However, the heat due to
layer accretion should reduce the temperature gradient
of the `dead zone' in the vertical direction, possibly 
leading to the suppression of weak convection as a result of
enhanced radiative losses in the radial direction (GVC). 
Therefore, the question becomes, does strong
convection transport angular momentum inward or outward?
Low Reynolds number
simulation by \cite{cp92} 
found strong convection transports angular 
momentum outward.

Azimuthal pressure perturbations have drawn large
attention in the literature recently,
as suggested by the equations of mean-square
velocity perturbation and by, in this paper,
the equations of enstrophy.
Axisymmetry/nonaxisymmetry of convection is actually
an intimate issue of the damping of convection due to
MHD turbulent mixing (\cite{khk99}).
When MHD turbulent mixing
is large ({\ie large $\alpha_{SS}$}), modes with large
$k_r$ are all suppressed owing to strong radial mixing.
Similarly, modes with large $k_r$ should also succumb to
strong radiative losses in radial direction.
Modes with small $k_r$ can
survive in a shearing environment only if $k_{\theta}$
is also very small, suggesting nearly axisymmetrical convection.
As mentioned in the preceding paragraph, weak convection
occurs when $\alpha_{SS}$ is large. Nearly axisymmetrical convection
is therefore a manifestation of weak convection which
struggles with turbulent and radiative damping in a
differentially rotating flow. In other words,
nonaxisymmetry of strong ($Ro \sim 1$) convection
means that
azimuthal pressure perturbations could be important to
violation of conservation of angular momentum. If this
happens, the equations of velocity and vorticity fluctuations
could allow strong convection to transport angular
momentum radially outward. 
Without a doubt, the quasi-linear
analysis based on equations of velocity and vorticity fluctuations
is very suggestive.


The other point of view regarding the
nature of positive/negative viscosity is the
direction of turbulent energy cascade. Negative
eddy viscosity manifests the process in which turbulent
fields do not only extract energy from the mean flow, but
the extracted energy is also passed up to larger scales and
finally goes back to the mean flow (\cite{starr68}).
The connection between
negative viscosity and inverse energy cascade in
two-dimensional flows has been studied in fluid
society (\cite{krai76}; \cite{pou78}; \cite{che99}).
Cabot 1996 and Klahr \etal
suggested that negative viscosity is a
result of an inverse energy cascade of nearly axisymmetrical 
convection and that positive viscosity results from
nonaxisymmetrical convection in which 3D hydrodynamical
turbulence is resumed. In fact, the collapse of 3D
hydrodynamical turbulence to 2D is not necessarily
related only to axisymmetrical patterns, but is 
a natural result
of turbulent anisotropy due to strong body forces,
which has been manifested by ample examples in geophysics.
Hossain 1994 found that when a
strong rigid rotation ($Ro<1$) is turned on,
the turbulent velocity
fields perpendicular to
the direction of rotation ({\it i.e.} $x-y$ plane)
are strongly correlated along
the direction of rotation ({\it i.e.} $z$ direction).
Therefore 3D
turbulence (direct cascade) reduces to an approximate
2D state (inverse cascade). In the case of accretion disks,
epicyclic effect should enforce turbulent fields of weak convection
to have an approximate 2D state, giving rise to negative
viscosity due to an inverse energy cascade. If convection is
strong, however, the collapse to 2D breaks down and this
probably results in positive/zero eddy viscosity due to a direct
energy cascade.

The numerical simulation by
\cite{kla00} found that the gradient of fluctuating angular
momentum generated by weak convection could flatten background
gradient of angular momentum. Consequently, a secondary instability
occurs at a fast rate of some fraction of $\Omega$. This induced
instability overwhelms weak convection and
drives angular momentum outward. In terms of the quasi-linear
analysis based on velocity and vorticity fluctuations, outward
angular momentum transport by this 
secondary instability is a result of strong azimuthal pressure
perturbations and stochastic stretching of vortex by turbulent
shear. By virtue of nonlinear phenomena of turbulent cascade, 
this fast secondary instability diminishes epicyclic effect
and thereby forces fluid back to an approximate
3-D hydrodynamical state.

If thermal convection generates positive eddy viscosity in
Keplerian disks, it does not necessarily
mean that it can be self-sustained. It has been a concern that
a self-sustained
convection would violate the second law of thermodynamics: in
the case of
a system in which the eddies are thermally driven, the heat
dissipated from the mean flow cannot again be used to drive
the eddies (\cite{starr68}). 
If thermal convection can be self-maintained in Keplerian disks, 
this
could mean that convection is not {\it totally} thermally
driven, in the sense that it can access the available rotational
energy at smaller radii via secondary or nonlinear instabilities
(\cite{knl95}; \cite{rz99}). However, the reverse of the above
statement is not necessarily true. \cite{kla00} shows that
circumstellar disks with outward transport of angular momentum
by convection cool continuously throughout the simulation.

Direction of angular momentum transport by convection is
also an important issue
in advection-dominated accretion flows (ADAFs).
When $\alpha_{SS}$ (not due to convection) is small,
strong convection ($Ro \sim 1$)
occurs when convection transports
angular momentum inward (or outward less efficiently than
it transports energy), and accretion
is suppressed in ADAFs (\cite{nia}; \cite{qg}).
However, as noted in this section, strong convection is unlikely
to be nearly axisymmetrical and is doubtful to be able to
collapse to a
2D state especially in a thick disk where the direction of
strong entropy gradient has a component along angular
momentum gradient.
Numerical observations of a transition of direction
of angular momentum (if it exists), a transition of
direction of energy cascade, and a transition of pattern
formation for convection in both Keplerian
accretion disks and ADAFs
is worthwhile when $Ro$ and $Ra$
are the control parameters, where $Ra$ should be the ``turbulent''
Rayleigh number characterized by radiative/advection losses and
MHD turbulent damping.

While the {\it perturbed}
baroclinic terms in the equations of perturbed enstrophy
act as the energy source of fluctuating vorticities,
we do not consider the terms associated with the
{\it background} baroclinic term in this paper,
due to the thin-disk approximation. 
That is, $\Omega$ does not vary with $z$ dramatically
in thin disks, resulting in a barotropic and Keplerian
disk (\eg see \cite{fkr92}).  The background baroclinic
effect is usually of less importance on local faster
turbulence in disks; for instance, the Biermann battery
owing to the same mechanism is usually ignored when
a mean-field dynamo is able to operate via
mean kinetic helicities driven by local turbulent events
such as thermal convection in stars
or supernovae/stellar winds
in galactic disks. In spite of their appearance
in thick disks which behave as fast rotating stars, 
it is also quite doubtful that the background baroclinic term 
can have important influence on the onset of convective instability
in these cases. 
As long as the convective growth rate is not too small, the effect
of large scale circulation due to the baroclinic term
to smooth away entropy inhomogeneity
associated with small-scale mixing such as thermal
convection in this case is usually negligible,
compared with the same damping mechanism achieved by MHD
turbulence. On the other hand, as we mention in the end of
\S 2, a large-scale vortex can be initiated from
the background baroclinic effect. Large-scale
eddies could be also formed via merging convective vortexes
(\cite{zes93}; \cite{gl99})
or through helicity fluctuations of convection (\cite{begm}).
The nonlinear interaction
between large-scale vortexes and convective vorticities
is worth investigating.

Finally we note that our analysis in this paper is applied
to local turbulent mixing in accretion disks, such as
thermal convection and the Balbus-Hawley instability.
Turbulent transport mediated by global waves, such as
internal, Rossby, or spiral shock waves, is not subject
to our analysis. Moreover, convective turbulence might
behave differently in the environment of
MHD turbulence since
ideal invariants during turbulent cascades
are different between hydrodynamical and MHD turbulence
(\eg see \cite{bis97}).

\acknowledgements
We are deeply indebted to Ethan T. Vishniac
for many useful discussions, and
for providing the information about Klahr 2000.
We are also grateful to 
the anonymous referee for insightful
comments and suggestions. This work was completed
in the High Energy Physics Laboratory at The
University of Texas at Austin, and we would like to
thank Roy Schwitters for his generous hospitality.


\end{document}